\documentclass[10pt,final,onecolumn]{IEEEtran}
\usepackage{amsmath,amsfonts,amssymb,color,psfrag,graphicx,subfigure}

\begin{document}
\title{Efficient Reconciliation of Correlated Continuous Random Variables using LDPC Codes}
\author{	Matthieu~Bloch$^{1}$, Andrew~Thangaraj$^{2}$, and Steven~W.~McLaughlin$^{1}$
					\thanks{$^{1}$GTL-CNRS Telecom, 2-3 rue Marconi 57070 Metz, France.}%
					\thanks{$^{2}$Dept of Electrical Engineering IIT Madras Chennai 600036, India.}%
}
\maketitle 
\begin{abstract}
This paper investigates an efficient and practical information reconciliation method in the case where two parties have access to correlated continuous random variables. We show that reconciliation is a special case of channel coding and that existing coded modulation techniques can be adapted for reconciliation. We describe an explicit reconciliation method based on LDPC codes in the case of correlated Gaussian variables. We believe that the proposed method can improve the efficiency of quantum key distribution protocols based on continuous-spectrum quantum states.
\end{abstract}
\IEEEpeerreviewmaketitle

\section{Introduction}
We consider here the situation where two parties (Alice and Bob) have access to correlated but non-identical random variables and wish to agree on a common bit sequence. The imperfect correlation of the random variables introduces discrepancies between their data which cannot be corrected without additional communication. Reconciliation is the process of finding and correcting these discrepancies~\cite{Brassard1993}. This problem can be viewed as a special case of source coding with side information~\cite{Slepian1973,Cover1991}; however the communication between the two parties is not bound to be one-way and interactive protocols could be used as well. An important application of reconciliation is the case where a third party (Eve) also has access to a correlated random variable but should have zero knowledge about Alice and Bob's reconciliated sequence. In fact this secret key agreement is usually performed in two steps: Alice and Bob first reconcile their data in order to get on a common sequence and then distill a secret key using privacy amplification techniques~\cite{Bennett1995}. The number of secret bits extracted by privacy amplification depends not only on the initial correlation between the random variables but also on the information leaked during reconciliation. Therefore one also often requires the reconciliation step to minimize the information leaked to the eavesdropper. 

The need for good reconciliation methods has recently appeared with the advent of \underline{Q}uantum \underline{K}ey \underline{D}istribution (QKD). In fact QKD offers practical means of providing Alice and Bob with correlated random variables while bounding the information available to Eve~\cite{Bennett1992,Gisin2002}. Let us briefly describe how secret key agreement is performed in such schemes. Alice first sends random quantum states to Bob via an insecure quantum channel. Noise in the channel introduces unavoidable errors in Bob's measures, however the laws of quantum mechanics also guarantee that any attempt to eavesdrop the channel introduces additional discrepancy between Alice and Bob's data. Simple bit-error-rate evaluations are then performed on a fraction of the data to upper-bound the information available to the eavesdropper. Alice and Bob correct the errors in the remaining part of their data by running a reconciliation protocol over a classical authenticated public channel, which allows them to agree on an identical sequence while minimizing the information leaked to the eavesdropper. Finally based on their estimation of the total information accessible to the eavesdropper, Alice and Bob extract a secret key using a privacy amplification protocol. The secret key is usually later used for cryptographic purposes, for instance to transmit secret messages with a one-time pad ~\cite{Shannon1948}.  

The reconciliation of discrete random variables has been extensively studied~\cite{Brassard1993,Muramatsu2003} and many practical and efficient interactive protocols (Cascade, Winnow) have been designed and are now widely used in QKD applications. However little work has been devoted to the reconciliation of continuous random variables. Such correlations appear for instance during some QKD protocols based on the continuous modulation of quantum Gaussian states~\cite{Cerf2001,Grosshans2002,Grosshans2003} and require specific reconciliation techniques. To our knowledge \underline{S}liced \underline{E}rror \underline{C}orrection (SEC)~\cite{VanAssche2004,Nguyen2004} is the only reconciliation method for continuous random variables proposed so far. SEC makes use of asymptotically efficient interactive error correcting codes however its efficiency in practical cases is still far away from its optimal bound. 

This paper investigates a new one-way reconciliation method inspired from coded modulation techniques with LDPC codes whose results slightly improve those of~\cite{VanAssche2004}. The remainder of the paper is organized as follows. Section~\ref{sec:ProblemDescription} shows how reconciliation can be viewed as a special case of channel coding with side information. In Section~\ref{sec:CodeDesign} we present a general practical reconciliation method and show its connection with the technique of SEC. Section~\ref{sec:Reconciliation} gives an explicit code construction and numerical results in the case of correlated Gaussian random variables, with code choice optimized using \underline{Ex}trinsic \underline{I}nformation \underline{T}ransfer (EXIT) charts.

\section{Reconciliation and error control coding}
\label{sec:ProblemDescription}
\subsection{Source coding with side-information}
\begin{figure}[htbp]
	\centering
		\includegraphics[width=0.50\linewidth]{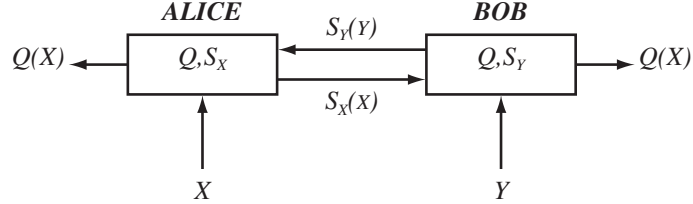}
	\caption{General reconciliation scheme.}
	\label{fig:Reconciliation}
\end{figure}

The two parties Alice and Bob each have access to the outcomes $\{x_i\}_{1..n}\in\mathbb{R}^n$ and $\{y_i\}_{1..n}\in\mathbb{R}^n$ of $n$ i.i.d. instances of two distinct correlated continuous random variables $X,Y\in\mathbb{R}$ with joint probability distribution $p(x,y)$. Alice and Bob then wish to distill a common binary string by exchanging information as shown in~Fig.~\ref{fig:Reconciliation}. We can assume without any restriction that the common binary string is the binary description of a quantized version of Alice's data only: $\{\mathcal{Q}(x_i)\}_{1..n}$ where $\mathcal{Q}$ denotes a quantizer. As already noticed in~\cite{VanAssche2004}, the quantization of $X$ into $\hat{X}=\mathcal{Q}(X)$ does not limit in itself the efficiency of the procedure since $I(\hat{X};Y)$ can be made arbitrarily close to $I(X;Y)$ by choosing a finer quantization. Note also that the discrete output of such a reconciliation procedure can then be used directly in conjunction with existing privacy amplification techniques.

As stated earlier reconciliation can be viewed as a special case of source coding with side information. In fact Alice has to send Bob a compressed version of her symbols $\{x_i\}_{1..n}$ taking into account the fact that he has access to the symbols $\{y_i\}_{1..n}$ as side information. If we let $I_{rec}^{min}$ be the minimum number of bits exchanged per symbol during reconciliation, we have (by the Slepian-Wolf theorem~\cite{Slepian1973}):
\begin{equation}
	I_{rec}^{min} = H(\hat{X}|Y).
\end{equation}
Note that the Slepian-Wolf theorem only applies to discrete random variables. However if we use a quantized version $\hat{Y}$ of $Y$ then the minimum information needed by Bob is $H(\hat{X}|\hat{Y})$ which can approach $H(\hat{X}|Y)$ with arbitrary precision~\cite{Cover1991}.

\subsection{Channel coding with side information}
The joint probability $p(x,y)$ of the continuous random variables $X$ and $Y$ can always be written as the product $p(y|x)p(x)$. 
In other words the symbols $\{y_i\}_{1..n}$ could have been obtained as the output of a memoryless channel $C_1$ characterized by the transition probability $p_1(y|x) = p(y|x)$ when the i.i.d symbols $\{x_i\}_{1..n}$ are present at the input. 

Let us now give a general description of the quantizer. Let $\{I_j\}_{1..k}$ be a partition of $\mathbb{R}$, let $\{\hat{x}_i\}_{1..k}\in{\{I_j\}_{1..k}}$ be the corresponding quantized values and let $\chi_j(x):\mathbb{R}\rightarrow\{0,1\}$ be the indicator function of interval $I_j$. The function $\mathcal{Q}:\mathbb{R}\rightarrow\mathcal{D}$ maps the elements of $\mathbb{R}$ to elements in the discrete set ${\mathcal{D}}=\{\hat{x_i}\}_{1..k}$ according to $\mathcal{Q}(x)=\sum_{j=1}^k{\hat{x}_j\chi_j(x)}$. Then the random variable $\hat{X}$ takes the discrete values $\{\hat{x}_j\}_{1..k}$ with probability $p_j=\mbox{Pr}\left[\hat{X}=\hat{x}_j\right]=\int{p(x)\chi_j(x)}dx$. Hence the symbols $\{x_i\}_{1..n}$ can also be viewed as the output of a discrete input/continuous output channel $C_2$ characterized by the transition probabilities $p_2(x|\hat{x}_j)=p(x)\chi_j(x)/p_j$ when the symbols $\{\mathcal{Q}(x_i)\}_{1..n}$ are present at the input.

\begin{figure}[htbp]
	\centering
		\includegraphics[width=0.50\linewidth]{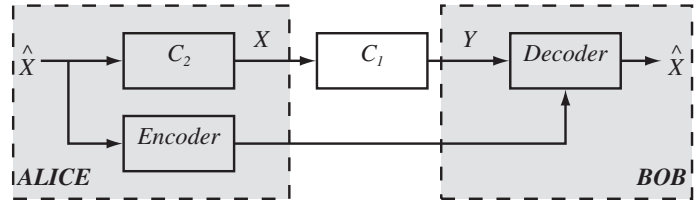}
	\caption{Reconciliation as channel coding.}
	\label{fig:Channel}
\end{figure}
	
Finally the continuous symbols $\{y_i\}_{1..n}$ could have been generated by sending the discrete symbols $\{\mathcal{Q}(x_i)\}_{1..n}$ through a channel $C_3$ obtained by concatenating channels $C_1$ and $C_2$. Since $\hat{X}\rightarrow X\rightarrow Y$ is a Markov chain the transition probability is given by $p(y|\hat{x}_j)=\int{p_1(y|x)p_2(x|\hat{x}_j)dx}$. The reconciliation of $X$ and $Y$ is then a special case of channel coding where symbols are transmitted over the channel $C_3$, while the error correcting information is transmitted separately over an error-free channel and available at the receiver as side information. Since the achievable code rates over this channel are upper bounded by $I(\hat{X};Y)\leq I(X;Y)$ we define the efficiency of a code with rate $R_c$ as:
\begin{equation}
	\eta = \frac{R_c}{I(X;Y)}=\frac{H(\hat{X})-I_{red}}{I(X;Y)},
\end{equation} 
where $I_{red}$ is the number of redundant information bits per symbol added by the code. A code achieving capacity will have an efficiency $\eta=I(\hat{X};Y)/I(X;Y)$ and will introduce $I_{red}^{min}$ bits of information per symbol:
\begin{equation}
	I_{red}^{min}=H(\hat{X})-I(\hat{X};Y)=H(\hat{X}|Y)
\end{equation}
Notice that $I_{red}^{min}$ and $I_{rec}^{min}$ are identical, hence this channel coding approach is strictly equivalent to the original source coding problem.

\section{Coded modulation for efficient reconciliation}
\label{sec:CodeDesign}
The practical efficiency of reconciliation relies on our ability to design good codes and decoders operating at a rate close to $I(\hat{X};Y)$. Since each quantized value $\hat{x}_j$ can be uniquely described by a $l$-bits label ($l=\left\lceil \log_2k\right\rceil$) we can define $l$ labeling functions $\mathcal{L}_m:\mathbb{R}\rightarrow\{0,1\}$ ($1\leq m\leq m$) that map any element $x\in\mathbb{R}$ to the $m$th bit of the label of $\mathcal{Q}(x)$. We then use the syndromes of $\{\mathcal{L}_m(x_i)\}_{1..l,1..n}$ according to a binary code as the side information sent by Alice to Bob on the public channel. Most standard coded modulation techniques such as \underline{B}it \underline{I}nterleaved \underline{C}oded \underline{M}odulation (BICM)~\cite{Caire1998,tenBrink1998} and \underline{M}ulti\underline{L}evel \underline{C}oding / \underline{M}ulti\underline{S}tage \underline{D}ecoding (MLC/MSD)~\cite{Wachsmann1999,Worz1992} schemes can be adapted and applied to reconciliation. Turbo codes and LDPC codes have already proved their excellent performance for error correction and side-information coding~\cite{Liveris2002} therefore both are good candidates for efficient reconciliation. This paper focuses on LDPC codes although we believe that turbo-codes or any other strong channel codes would yield similar results.

\subsection{BICM and MLC/MSD}
Let $\{b_{i,1}\dots b_{i,l}\}$ denotes the $l$ label digits of Alice's quantized number $\mathcal{Q}(x_i)$. A soft-information decoder in a BICM or MLC/MSD scheme processes the data $\{y_i\}_{1..n}$ and computes the \underline{L}og-\underline{L}ikelihood \underline{R}atios (LLR) $\{\lambda_{i,i'}\}_{1..n,1..l}$:
\begin{equation}
\lambda_{i,i'}=\log\frac{\mbox{Pr}\left[b_{i,i'}=1|\{y_k\}_{1..n}\right]}{\mbox{Pr}\left[b_{i,i'}=0|\{y_k\}_{1..n}\right]}.
\end{equation}
This LLR can be written as the sum of two distinct contributions $\lambda_{i,i'}^{int}$ and $\lambda_{i,i'}^{ext}$ which respectively describe intrinsic and extrinsic information:
\begin{eqnarray}
\lambda_{i,i'}^{ext}&=&\log\frac{\mbox{Pr}\left[b_{i,i'}=1|\left\{y_k\right\}_{k\neq i}\right]}{\mbox{Pr}\left[b_{i,i'}=0|\left\{y_k\right\}_{k\neq i}\right]},\label{eq:intrinsic}\\
\lambda_{i,i'}^{int}&=&\log \frac{\sum_{\hat{x}_j\in\mathcal{A}_{i',1}}\left[\prod_{m\neq i'}\mbox{Pr}\left[b_{i,m}=\mathcal{L}_m(\hat{x}_j)|\left\{y_k\right\}_{k\neq i}\right]\int{p(y_i,x)\chi_j(x)dx}\right]}{\sum_{\hat{x}_j\in\mathcal{A}_{i',0}}\left[\prod_{m\neq i'}\mbox{Pr}\left[b_{i,m}=\mathcal{L}_m(\hat{x}_j)|\left\{y_k\right\}_{k\neq i}\right]\int{p(y_i,x)\chi_j(x)dx}\right]},\label{eq:extrinsic}
\end{eqnarray}
with $\mathcal{A}_{i',b}=\{\hat{x}_j\in\mathcal{D}:\mathcal{L}_{i'}(\hat{x}_j)=b\}$ for $b\in\{0,1\}$. The intrinsic part $\lambda_{i,i'}^{int}$ exploits the existing correlation between $X$ and $Y$ to gather information about $b_{i,i'}$. Each term $\mbox{Pr}\left[b_{i,m}=\mathcal{L}_m(\hat{x}_j)|\left\{y_k\right\}_{k\neq i}\right]$ is actually a function of $\lambda_{i,m}^{ext}$ and $\mathcal{L}_m(\hat{x}_j)$, therefore $\lambda^{int}_{i,i'}$ also gathers all the information about $b_{i,i'}$ contained in the previously decoded bits $\{b_{i,m}\}_{m\neq i'}$. The expressions~(\ref{eq:intrinsic}) and~(\ref{eq:extrinsic}) are then ideally suited for iterative calculations.

\begin{figure}[htbp]
	\centering
		\includegraphics[width=0.50\linewidth]{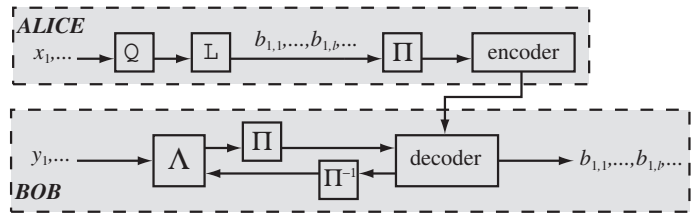}
	\caption{BICM-like reconciliation. Alice's continuous data is sent through a quantizer $\mathcal{Q}$, a binary mapper $\mathcal{L}$ and an interleaver $\Pi$. The encoder computes syndromes which are then directly available on Bob'side. Bob decodes his data by iteratively demapping his symbols ($\Lambda$-box) and correcting errors.}
	\label{fig:BICM}
\end{figure}

\begin{figure}[htbp]
	\centering
		\includegraphics[width=0.50\linewidth]{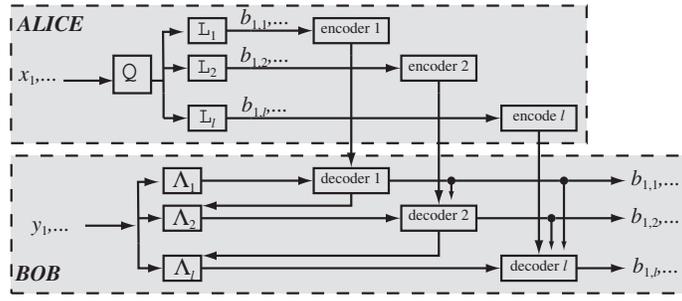}
	\caption{MLC/MSD-like reconciliation. Alice's data is sent through a quantizer $\mathcal{Q}$ and $l$ distinct binary mappers $\mathcal{L}_m$. Syndromes are computed separately for each level. Bob decodes his symbols by iteratively processing the $l$ levels.}
	\label{fig:MLCMSD}
\end{figure}

In the case of a BICM-like reconciliation a single code would be applied to an interleaved version of the whole bit string $\{\mathcal{L}_m(x_i)\}_{1..l,1..n}$, see~Fig.~\ref{fig:BICM},whereas in the case of MLC/MSD-like reconciliation $l$ individual codes would be applied  successively to the $l$ binary sequences $\mathcal{L}_m({\{x_i\}_{1..n}})$, see~Fig.~\ref{fig:MLCMSD}. Notice that when capacity approaching codes are used MLC/MSD is optimal whereas BICM in only suboptimal. MLC/MSD also usually offers more flexibility than BICM. There is no general method for designing efficient capacity approaching codes over the equivalent channel $C_3$ of section~\ref{sec:ProblemDescription}. However when the correlation between $X$ and $Y$ is symmetric ($p(y,x)=p(-y,-x)$) and when the chosen quantizer and labeling strategy also preserve this symmetry property one could use LDPC codes optimized by density evolution~\cite{Richardson2001a}.

\subsection{SEC as multistage hard decoding}
Interestingly the original SEC protocol is just a special case of MLC/MSD-like reconciliation where interactive \underline{B}inary \underline{C}orrection \underline{P}rotocols (BCP) optimized for BSC are used as component codes and where the information passed between levels is simply the sign of $\lambda_{i,i'}^{int}$. Since these BCPs perform perfect error correction it is sufficient to process the $l$ levels successively.

Let us briefly analyze the cost of this operation compared to optimal MLC-MSD. Let $\{\mathcal{L}_1(\hat{X})\dots\mathcal{L}_l(\hat{X})\}$ be the $l$ random variables corresponding to the $l$ label bits of $\hat{X}$. Then using the chain rule of mutual information:
\begin{eqnarray}
	I(\hat{X};Y) &=& I(\mathcal{L}_1(\hat{X})\dots\mathcal{L}_l(\hat{X});Y) \nonumber\\
	&=& \sum_{i=1}^{l}{I(\mathcal{L}_i(\hat{X});Y|\mathcal{L}_1(\hat{X})\dots\mathcal{L}_{i-1}(\hat{X}))}\nonumber\\
     &=&\sum_{i=1}^{l}{\left[H(\mathcal{L}_i(\hat{X})|\mathcal{L}_1(\hat{X})\dots\mathcal{L}_{i-1}(\hat{X}))-H({\mathcal{L}_i(\hat{X})|Y,\mathcal{L}_1(\hat{X})\dots\mathcal{L}_{i-1}(\hat{X})})\right]}
\end{eqnarray}
Let $\mathcal{M}_i\left({Y,\mathcal{L}_1(\hat{X})\dots\mathcal{L}_{i-1}(\hat{X}))}\right)$ be the maximum-likelihood estimation of $\mathcal{L}_i(\hat{X})$ used in SEC. If we assume that the symmetry condition is valid then $H(\mathcal{L}_i(\hat{X})|\mathcal{L}_1(\hat{X})\dots\mathcal{L}_{i-1}(\hat{X}))=1$. and using Fano's inequality:
\begin{eqnarray}
	I(\hat{X};Y) 
	\geq \sum_{i=1}^{l}\left[1-h(p_i)\right]
\end{eqnarray}
where $p_i=\mbox{Pr}\left[\mathcal{L}_i(\hat{X})\neq \mathcal{M}_i\left({Y,\mathcal{L}_1(\hat{X})\dots\mathcal{L}_{i-1}(\hat{X}))}\right)\right]$ and $h$ is the binary entropy function. As expected treating all $l$ levels as BSCs underestimates $I(\hat{X};Y)$ and $\sum_{i=1}^{l}\left[1-h(p_i)\right]
$ only heads toward $I(X;Y)$ asymptotically as $l\rightarrow\infty$. For practical values of $l$ (say less than 5) this approximation may not be tight enough to ensure a good reconciliation efficiency, even with perfect codes achieving capacity over the BSC. Note also that a careful evaluation of the information leaked is needed when interactive BCPs are used. The most general evaluation should take into account all the bits exchanged in both ways which dramatically reduces the efficiency of error correction at levels where the bit error rate is above a few percents. 

Despite its simplicity the practical interest of SEC with interactive BCPs is threefold. As stated earlier BCPs allow the two parties to achieve perfect error correction whereas one-way code only correct errors approximately (for instance the sum-product decoding of LDPC codes may leave a couple of errors uncorrected). The second advantage is that one can blindly apply an interactive BCP without bothering about the actual code rate needed while one-way code require a specific design for each rate. Finally the decoding complexity of BCPs is very low when compared to belief-propagation decoding .

\section{Reconciliation of Gaussian Variables}
\label{sec:Reconciliation}
In this section we deal with the reconciliation of two Gaussian random variables $X\sim\mathcal{N}(0,\Sigma)$ and $Y=X+\epsilon$ where $\epsilon\sim\mathcal{N}(0,\sigma)$ using MLC/MSD-like and BICM-like reconciliation with LDPC codes. Such correlations appear for instance during Gaussian modulated QKD protocols.

Since the probability distribution $p(y,x)$ is Gaussian it satisfies the symmetry condition mentioned earlier. Following the quantization technique proposed in ~\cite{VanAssche2004} the set of real numbers was split into an even number $k$ of  intervals $\{I_j\}_{1..k}$ symmetric around $0$ (this ensures the symmetry of the joint pdf between $\hat{X}$ and $Y$). The interval bounds were optimized using the simplex method in order to maximize $I(\hat{X};Y)$. 

\subsection{Choice of codes and rates for MLC/MSD-like reconciliation}
\label{sec:MLCMSDlike}

\begin{figure}[htbp]
	\centering
		\subfigure[Binary mapping.\label{fig:Binary}]{\includegraphics[width=0.45\linewidth]{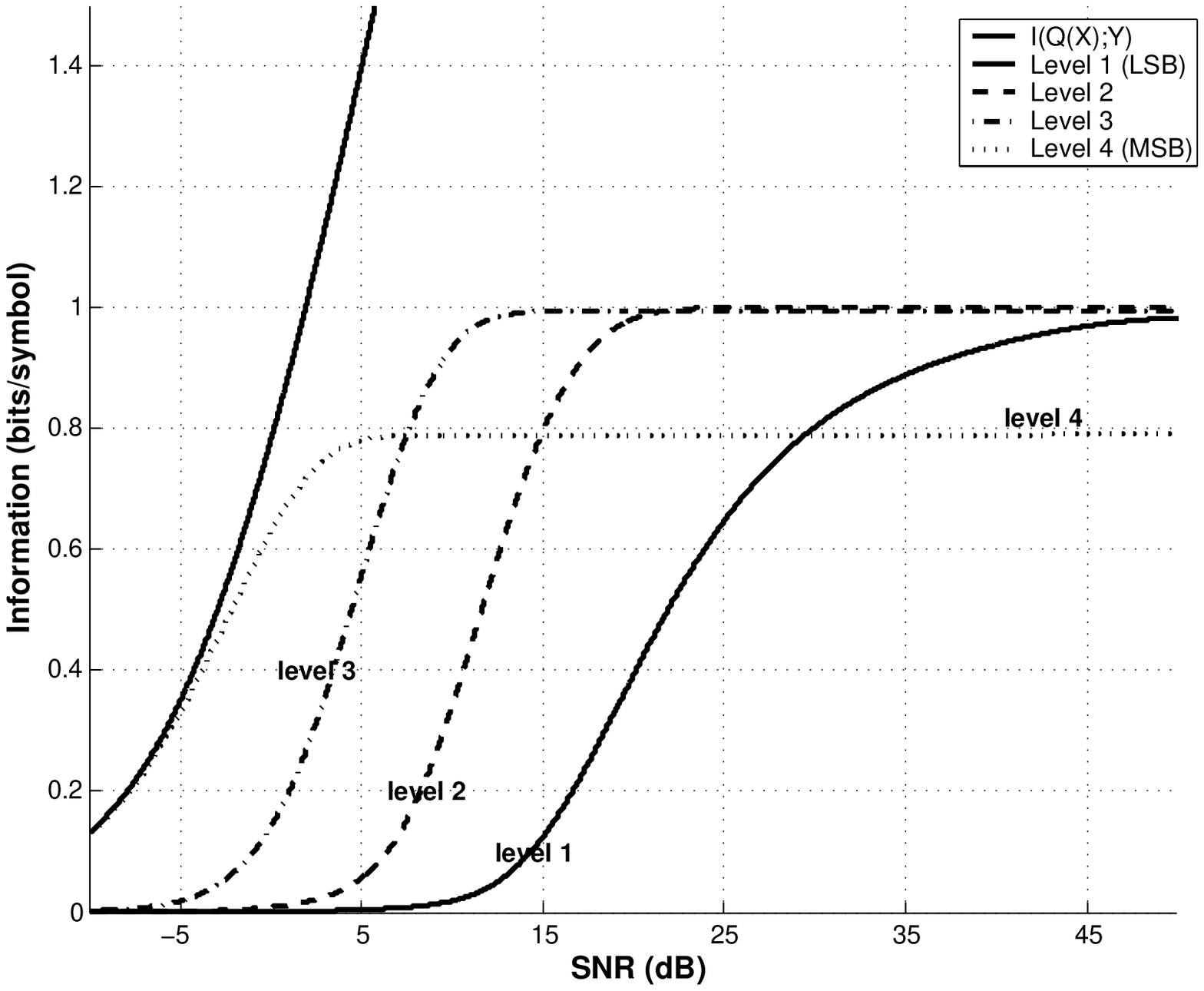}}
		\subfigure[Anti-binary mapping.\label{fig:Antibinary}]{\includegraphics[width=0.45\linewidth]{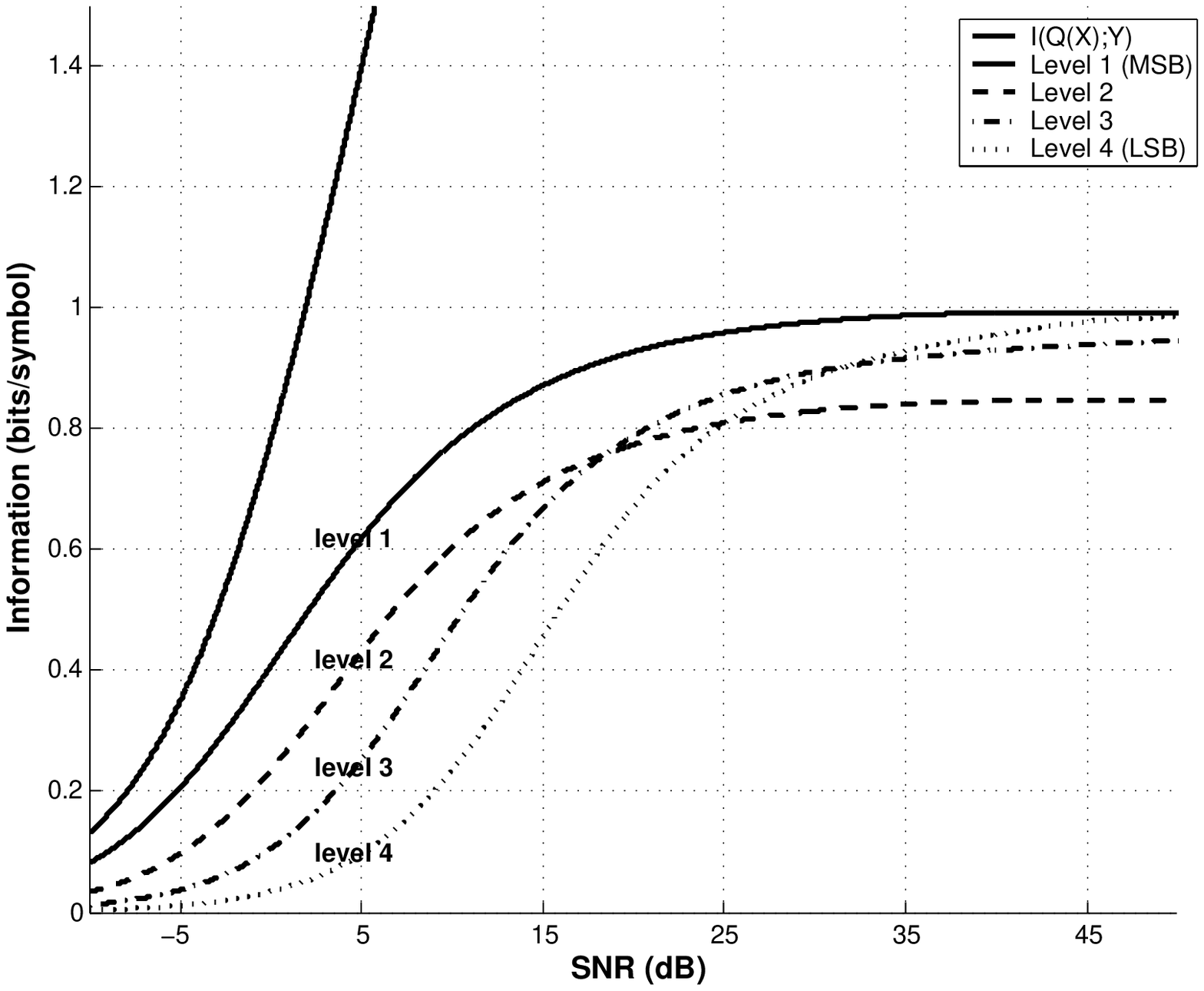}}
	\caption{Mutual information by level.}
\end{figure}

In order to preserve the symmetry property a labeling strategy for MLC/MSD-like reconciliation should satisfy: 
\begin{equation}
 \forall m\qquad p\left(y,\mathcal{L}_m(\mathcal{Q}(x))=b\right)=p\left(-y,\mathcal{L}_m(\mathcal{Q}(x))=\overline{b}\right)	
 \label{eq:symmprop}
\end{equation}
where $b\in\{0,1\}$ . We investigated two labeling strategies fulfilling this requirement: binary and anti-binary mappings. Both mappings assign to each interval $j$ the $l$-bit representation ($l=\left\lceil \log_2k \right\rceil$) of $j+(2^n-k)/2$ but in the binary case the least significant bit level is decoded first while in the antibinary case the most significant bit level is decoded first. The mutual informations $I(\mathcal{L}_i(\hat{X});Y|\mathcal{L}_1(\hat{X})\dots\mathcal{L}_{i-1}(\hat{X})$ in these two cases are plotted as a function of the normalized SNR $10\log(\Sigma^2/2\sigma^2)$ when $k=16$ $(l=4)$ on Fig.~\ref{fig:Binary} and Fig.~\ref{fig:Antibinary}. 

Note that for low SNRs ($\leq 2$dB) the first two levels in the binary case are extremely noisy and the mutual information is close to 0. Designing good LDPC codes at such low rates is extremely hard and it is easier to disclose the entire level without compromising efficiency. With antibinary mapping all levels have non-zero mutual-information and no such simplification can be made. Therefore in all MLC/MSD simulations we used a binary mapping, and by carefully choosing the number of quantization intervals the number of codes actually needed for reconciliation was easily reduced to 2 or 3.

The optimal code rate required at level $i$ with a given SNR is:
\begin{equation}
	R^i_{opt}=1-(I_i(\infty)-I_i(SNR)),	
\end{equation}
where $I_i(s)=I(\mathcal{L}_i(\hat{X});Y|\mathcal{L}_1(\hat{X})\dots\mathcal{L}_{i-1}(\hat{X}),SNR=s)$. For instance the rates required with 16 quantization intervals, binary mapping at an SNR $\Sigma^2/\sigma^2=3$ are 0.002/0.016/0.259/0.921. In this case the effect of quantization is negligible since $I(\hat{X};Y)$ differs from $I(X;Y)$ by less than $0.02$ bit/channel use. Full disclosure of the first two levels also has very little impact since these two levels contribute to $I(\hat{X};Y)$ by less than $0.02$ bit.

In order to further simplify the code design we used irregular LDPC codes optimized for the AWGN channel as component codes. Good degree distributions with threshold close to capacity can easily be obtained via density evolution~\cite{LTHC}. The block length used was $200,000$ and graphs were randomly generated while avoiding loops of length 2 and 4. Despite their long block length the performances of all constructed codes were still well below those of their capacity achieving  counterparts. Achieving perfect error correction with high probability is in fact only possible at the cost of reducing the code rates. Cutting down the rates of each component code would disclose far too many bits, but a careful choice of the code and iterations between levels make it possible to achieve good reconciliation efficiency. 

We investigated a code choice strategy based on EXIT charts~\cite{tenBrink2001}. EXIT charts are a convenient tool to visualize the exchange of mutual information between the decoders and the demappers involved in MLC/MSD as shown in~Fig.~\ref{fig:MLCMSD}. 

\begin{figure}[htbp]
	\centering
		\includegraphics[width=0.50\linewidth]{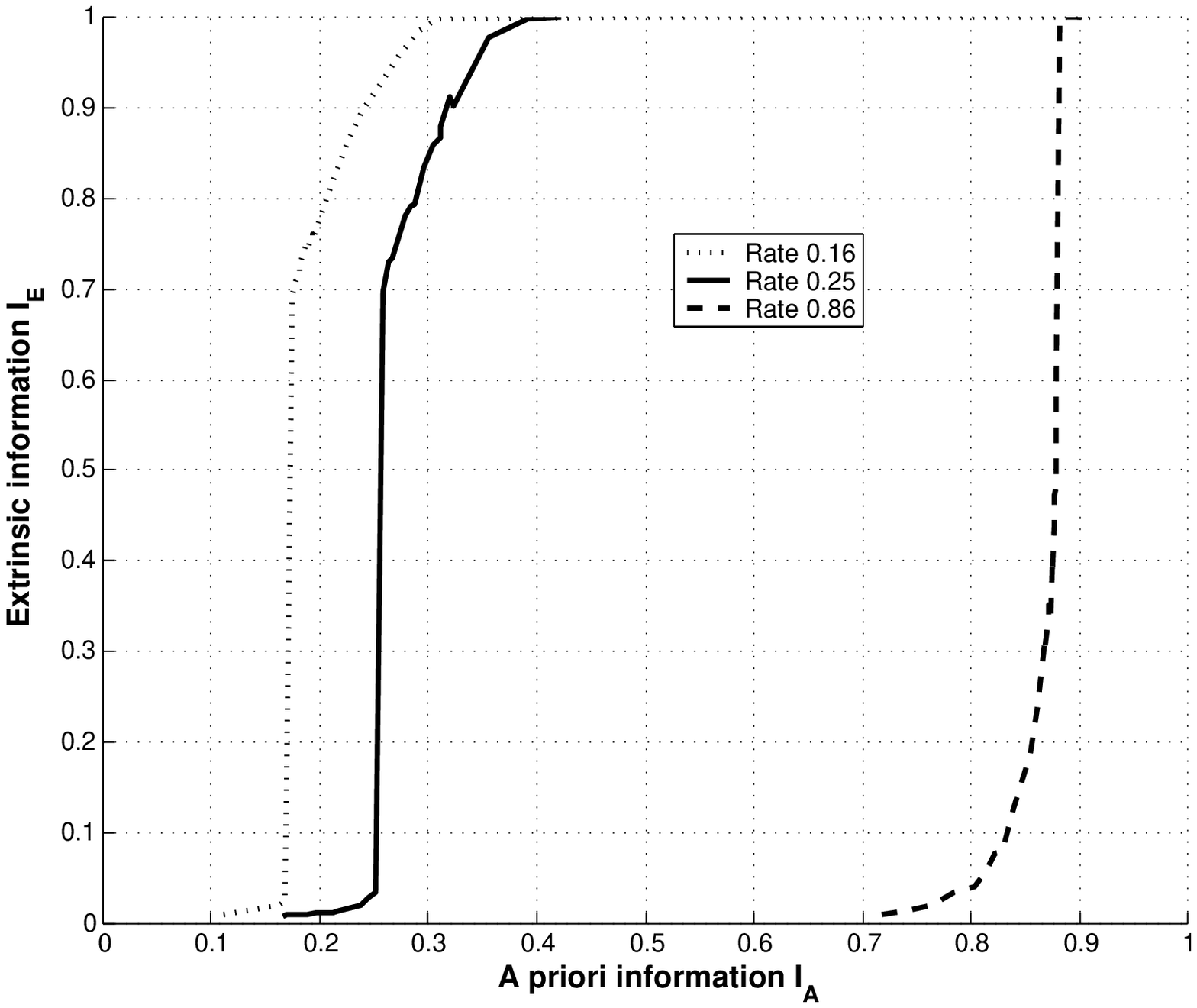}
	\caption{Transfer curves of constructed LDPC codes.}
	\label{fig:EXITcurve}
\end{figure}

The demapper transfer curves $I_E=T_d(I_A)$ cannot be computed in closed form but can be obtained via Monte-Carlo simulations using Eq.~(\ref{eq:extrinsic}). The transfer curves $I_E=T_c(I_A)$ of the constructed LDPC codes were also obtained by Monte-Carlo simulations with Gaussian \textit{a priori} information. The symmetry property allowed us to limit our simulations to the decoding of the all-zero codeword. Examples of transfer curves obtained with 100 iterations of sum-product decoding are given in Fig.~\ref{fig:EXITcurve}. As expected our simulations showed that low rate codes gathered far less extrinsic information than high rate codes. We decided to achieve perfect error correction by compromising on the rate of the high rate codes and by using iterations to compensate for the poor performance of lower rate codes.

\begin{figure}[htbp]
	\centering
		\includegraphics[width=0.50\linewidth]{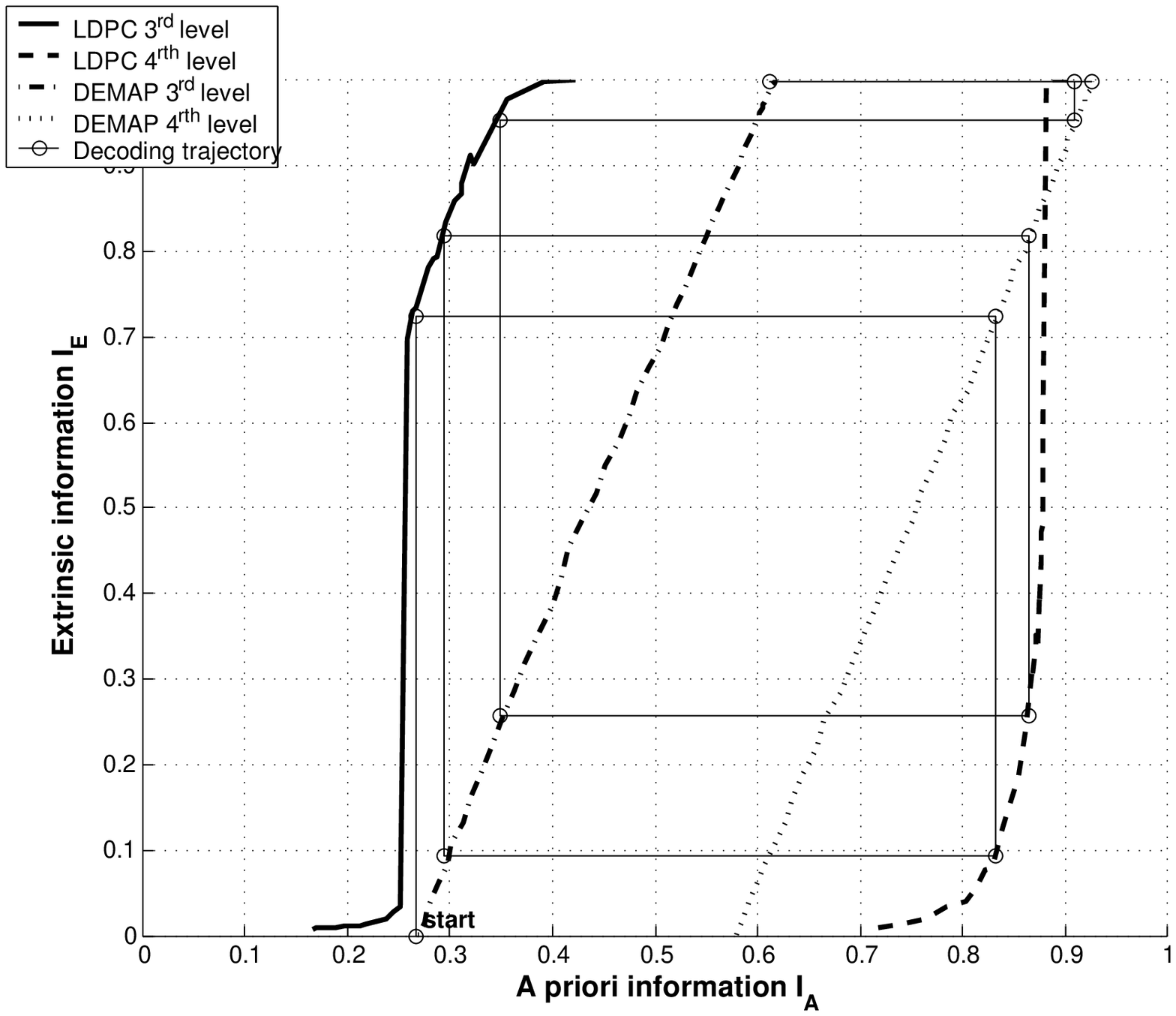}
	\caption{Iterative decoding trajectory when $\Sigma^2/\sigma^2=3$ with 16 quantization intervals and binary mapping. Decoding trajectory is averaged over 10 blocks.}
	\label{fig:EXITchart}
\end{figure}

Let us now detail how practical codes rates were found in the case $\Sigma^2/\sigma^2=3$ with 16 quantization intervals (4 bits) and a binary mapping. As explained above the first two levels are disclosed and one would in theory need two ideal codes with rate $0.26$ and $0.92$ to perform MSD. We used instead a rate $0.25$ code for the 3\textsuperscript{rd} level and looked for a high rate code that would gather enough extrinsic information at low SNR to start the decoding process and would correct all errors when a-priori information is $0.92$. This search was performed by drawing EXIT charts and ensuring that iterations would allow complete decoding. We found that a LDPC code with rate $0.86$ was a good compromise. Fig.~\ref{fig:EXITchart} shows that realistic decoding trajectories are closed to the expected decoding behavior.  
The code rates obtained with the same procedure for different values of SNR are given in Table~\ref{tab:ParamMLCMSD}.
\begin{table*}[htbp]
	\centering
		\begin{tabular}{|c|c|c|c|c|c|}
			\hline
			SNR	& Intervals				&$I(\hat{X};Y)$ &$H(\hat{X})$	&Optimum code rates									& Practical LDPC code rates \\
			\hline
			1		&	12 (4 levels)		&	0.49					& 3.38				& 0.001/0.008/0.187/0.915 	&	$0/0/0.16/0.86$					\\
			3		&	16 (4 levels)		& 0.98					&	3.78				& 0.002/0.016/0.259/0.921 	&	$0/0/0.25/0.86$					\\
			7		&	22 (5 levels)		& 1.47					& 4.23				& 0.002/0.020/0.295/0.924/1	&	$0/0/0.28/0.86/1$					\\
			15	&	32 (5 levels)		& 1.97					& 4.68				& 0.002/0.025/0.332/0.934/1 &	$0/0/0.31/0.86/1$					\\
			\hline
		\end{tabular}
	\caption{Parameters used for MLC/MSD-like reconciliation.}
	\label{tab:ParamMLCMSD}
\end{table*}
When rate-1 codes where required we used algebraic codes with error correcting capability of 1 instead of LDPC codes to correct the few erroneous blocks.

\subsection{Choice of codes and rates for BICM-like reconciliation}
\label{sec:BICMlike}

In the previous section we constrained the mapping to satisfy the symmetry condition~(\ref{eq:symmprop}) in order to simplify the code design at each level. In the case of BICM-like reconciliation a single code is now applied to an interleaved version of the label bits . Therefore provided that the mapping produces a balanced number of 0's and 1's a code optimized for symmetric mappings will behave almost equally well with non-symmetric ones. Note that with LDPC codes no additional interleaving is needed  due to the structure of the code. 

\begin{figure}[htbp]
	\centering
		\includegraphics[width=0.50\linewidth]{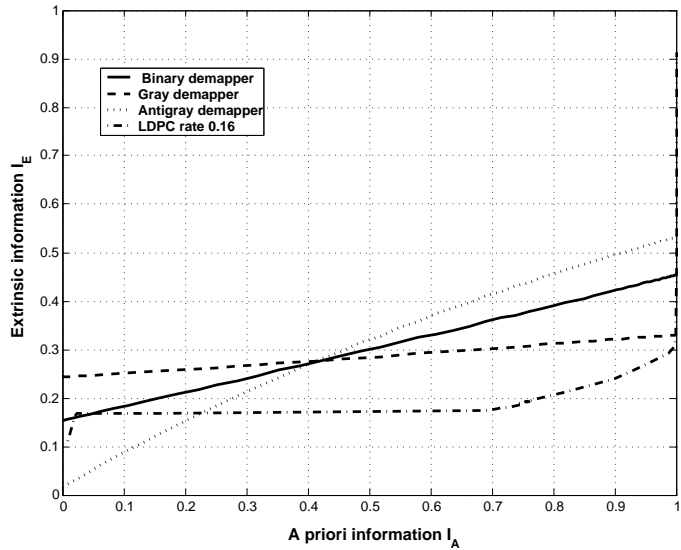}
	\caption{Transfer curves of demapper and code used in BICM-like reconciliation for $\Sigma^2/\sigma^2=3$ and 16 quantization intervals.}
	\label{fig:BICMcomp}
\end{figure}

The optimal code rate required with a given SNR is:
\begin{equation}
	R_{opt} = 1-\frac{H(\hat{X})-I(SNR)}{l},
\end{equation}
where $I(s)$ is the maximum BICM-capacity at SNR $s$. $I(s)$ depends on the mapping and cannot be computed exactly, however if we let $X_m\in{\{0,1\}}$ be the binary random variable at level $m$ ($1\leq m\leq l$) then we can estimate a lower and an upper bound:
\begin{equation}
	H(\hat{X})-\sum_{m=1}^{l}{H(X_m|Y)}\leq I(s) \leq \min\left\{H(\hat{X}),\sum_{m=1}^{l}I(X_m;Y)\right\},
\end{equation}
and obtain bounds on the optimal code rate:
\begin{equation}
	1-\frac{\sum_{m=1}^{l}H(X_m|Y)}{l}\leq R_{opt} \leq 1-\frac{\max\left\{0,H(\hat{X})-\sum_{m=1}^{l}I(X_m;Y)\right\}}{l}.
\end{equation}
For instance with 16 quantization intervals, $\Sigma^2/\sigma^2=3$ and a gray mapping the optimal rate is between $0.257$ and $0.274$. The maximum reconciliation efficiency is therefore less than 88\%. 
When choosing a code one has to ensure that the rate is also compatible with the mapping used. Fig.~\ref{fig:BICMcomp} shows the transfer curves of a rate 0.16 LDPC code optimized for the Gaussian channel as well as various demapper transfer curves. All transfer curves where obtained via Monte-Carlo simulations with Gaussian \textit{a priori} information. Perfect decoding is possible if the LDPC code transfer curve remains below the demapper curve. It clearly appears that all mappings cannot be used and that no mapping can gather high extrinsic information for both low and high \textit{a priori} information. Gray mapping gathers the highest extrinsic information without \textit{a priori} information but the slope of its transfer curve is the steepest, which means that it has to be associated with a strong code. The other mappings can be used with weaker but lower rate codes and are therefore not suitable for efficient reconciliation.

Unfortunately even with Gray mapping and a strong code we found that the practical code rates were far below the optimal ones. As shown in Fig.~\ref{fig:BICMcomp} a rate 0.16 LDPC code is required to ensure full error correction even though the demapper initially feeds the decoder with 0.24 \textit{a priori} information bits (note that this value is close to the lower bound on the optimal rate). 

\subsection{Numerical results}
Table~\ref{tab:ReconciliationEfficiency} shows the reconciliation efficiency obtained with our MLC/MSD-like procedure for different values of the SNR and compares it with the efficiency of SEC. Our simulations were performed over 50 blocks of size 200,000 and all errors were corrected. When rate-$1$ codes were required we used a BCH code with block length $4091$ and error correcting capability $t=1$. This disclosed slightly less than $0.003$ additional bits per symbol sent. Since high-rate LDPC codes would sometimes fail to correct a couple of erroneous bits we also applied the same BCH code on top of these LDPC codes. 

All SEC results are given for a quantization with 32 intervals. The efficiencies $\eta_{\mbox{\tiny SEC}}^{max}$, $\eta_{\mbox{\tiny SEC}}^1$ and $\eta_{\mbox{\tiny SEC}}^2$ respectively refer to the efficiency of SEC with ideal binary codes, interactive Cascade and one-way Cascade + Turbo-codes as reported in~\cite{VanAssche2005}. $\eta_{\mbox{\tiny MLC}}$ is the efficiency obtained with MLS/MSD like reconciliation using the code rates and quantizers of Table~\ref{tab:ParamMLCMSD}, while $\eta^{max}_{\mbox{\tiny MLC}}$ is the maximum efficiency attainable with capacity achieving codes.

\begin{table}[htbp]
	\centering
		\begin{tabular}{|c|c|c|c|c|c|}
			\hline
			SNR	& $\eta_{\mbox{\tiny SEC}}^{max}$& $\eta_{\mbox{\tiny SEC}}^1$ 	& $\eta_{\mbox{\tiny SEC}}^2$ 	& $\eta^{max}_{\mbox{\tiny MLC}}$& $\eta_{\mbox{\tiny MLC}}$	\\
			\hline
			1		&	75\%					&	60\%						&	$<$50\%					&	98\% 			&	79.4\%					\\
			3		&	87\%					&	79\%						&	67\%						&	98\% 			&	88.7\%					\\
			7		&	90\%					&	84\%						&	76\%						&	98\% 			&	90.9\%					\\
			15	&	92\%					&	87\%						&	82\%						&	98.5\% 		&	92.2\%				\\
			\hline
		\end{tabular}
	\caption{Reconciliation efficiency.}
	\label{tab:ReconciliationEfficiency}
\end{table}
The proposed reconciliation procedure achieves close if no better efficiency that SEC with ideal codes. When compared to SEC with one-way codes our procedure improves the efficiency significantly.
\section{Discussion and conclusion}
We have shown that reconciliation is equivalent to channel coding with side information and that existing coded-modulation methods such as MLC/MSD and BICM can be adapted. We noticed that SEC is in fact a sub-optimal implementation of MLC/MSD-like reconciliation. In the case of correlated Gaussian random variables we proposed a method to design efficient reconciliation schemes based on iterative MLC/MSD and BICM techniques with LDPC codes. Our simulations showed that MLC/MSD-like reconciliation achieves high efficiency and significantly improves previously reported results. 

Although the proposed reconciliation method performs well with LDPC codes optimized for the AWGN channel we believe that a more specific optimization would further improve the efficiency. For instance one could optimize LDPC codes with density evolution by taking into account the decoders and the demappers jointly. However from a practical prospective the procedure would lose its usefulness if the complexity of optimization procedure becomes significant. 

Finally our study was restricted to real random variables but the procedure can be straightforwardly extended to variables in $\mathbb{R}^n$. 

 
\bibliographystyle{IEEEtran}
\bibliography{MyBiblio}
 
\end{document}